\documentclass[runningheads]{llncs}

\RequirePackage{fix-cm}

\usepackage[noadjust]{cite}

\usepackage{graphicx}
\usepackage{epsfig}
\usepackage{dsfont}
\usepackage{amsmath}
\usepackage{amssymb}
\usepackage{boldline,multirow}
\usepackage{diagbox}
\usepackage{ulem}
\usepackage{color}
\usepackage{graphicx}
\usepackage{subfigure}
\usepackage{bbding}
\usepackage[colorlinks,
linkcolor=black,
anchorcolor=blue,
citecolor=blue
]{hyperref} 
\usepackage{makecell}
\makeatletter
\newcommand\figcaption{\def\@captype{figure}\caption}
\newcommand\tabcaption{\def\@captype{table}\caption}
\makeatother

\usepackage{array}
\makeatletter
\newcommand{\thickhline}{%
    \noalign {\ifnum 0=`}\fi \hrule height 1pt
    \futurelet \reserved@a \@xhline
}

\begin{document}
\title{Universal Segmentation of 33 Anatomies}% \textcolor{red}{with Cross-patch Transformer}}% \thanks{Supported by organization x.}}
%
%\titlerunning{Abbreviated paper title}
% If the paper title is too long for the running head, you can set
% an abbreviated paper title here
%
\author{Pengbo Liu\inst{2} \and
	Yang Deng\inst{1}\and
	Ce Wang\inst{2}\and
	Yuan Hui\inst{2}\and
	Qian Li\inst{2}\and
	Jun Li\inst{2}\and
	Shiwei Luo\inst{4}\and
	Mengke Sun\inst{2}\and
	Quan Quan\inst{2} \and
	Shuxin Yang\inst{2}\and
	You Hao\inst{2}\and
		Honghu Xiao\inst{3}\and
		Chunpeng Zhao\inst{3}\and
		Xinbao Wu\inst{3}\and
	S. Kevin Zhou\inst{1,2}}
%\author{First Author\inst{1}\orcidID{0000-1111-2222-3333} \and
%Second Author\inst{2,3}\orcidID{1111-2222-3333-4444} \and
%Third Author\inst{3}\orcidID{2222--3333-4444-5555}}
%
\authorrunning{Pengbo Liu et al.}
% First names are abbreviated in the running head.
% If there are more than two authors, 'et al.' is used.
%
\institute{
		School of Biomedical Engineering \& Suzhou Institute for Advanced Research Center for Medical Imaging, Robotics, and Analytic Computing \& Learning (MIRACLE) University of Science and Technology of China, Suzhou 215123, China \and
		Key Lab of Intelligent Information Processing of Chinese Academy of Sciences (CAS),
		Institute of Computing Technology, CAS, Beijing, 100190, China \email{liupengbo2019@ict.ac.cn}\and
		Beijing Jishuitan Hospital, Beijing 100035, China\and
		Department of Radiology, Guangzhou First People’s Hospital, School of Medicine, South China University of Technology, Guangzhou, China}

\maketitle              % typeset the header of the contribution
\begin{abstract}
In the paper, we present an approach for learning a single model that universally segments 33 anatomical structures, including vertebrae, pelvic bones, and abdominal organs. Our model building has to address the following challenges. Firstly, while it is ideal to learn such a model from a large-scale, fully-annotated dataset, it is practically hard to curate such a dataset. Thus, we resort to learn from a union of multiple datasets, with each dataset containing the images that are partially labeled. Secondly, along the line of partial labelling, we contribute an open-source, large-scale vertebra segmentation dataset  for the benefit of spine analysis community, \textbf{CTSpine1K}, boasting over 1,000 3D volumes and over 11K annotated vertebrae. Thirdly, in a 3D medical image segmentation task, due to the limitation of GPU memory, we always train a model using cropped patches as inputs instead a whole 3D volume, which limits the amount of contextual information to be learned. To this, we propose a \textbf{cross-patch transformer} module to fuse more information in adjacent patches, which enlarges the aggregated receptive field for improved segmentation performance. This is especially important for segmenting, say, the elongated spine. Based on 7 partially labeled datasets that collectively contain about 2,800 3D volumes, we successfully learn such a universal model. Finally, we evaluate the universal model on multiple open-source datasets, proving that our model has a good generalization performance and can potentially serve as a solid foundation for downstream tasks.

%segmentation model task with many specific categories. The most intuitive method is training \textit{many models} for different anatomies \textit{with corresponding partially labeled datasets}, which limits the \underline{scale} of dataset can be used to train our model. And the separated models are not efficient and convenient enough in clinical application. 

%We plan to open-source a \textit{large-scale} vertebrae segmentation dataset, CTSpine1K, with more than 1,000 3D volumes to enlarge the scale of datasets in bone segmentation community, and we trained a \textit{single} 3D universal segmentation model to segment 33 anatomies with only one-time inference via simple semi-supervised method based on 7 partially labeled datasets, including about 2,500 3D volumes.
%And in 3D medical image segmentation task, due to the limitation of GPU memory, we always trained the model using cropped patches as input instead the whole 3D volume, which also limits the amount of the information of the input. We proposed a \textit{cross-patch transformer module} to fuse more information in adjacent patches to enhance the `real' receptive field then the performance of the model, especially for segmenting anatomy with long distance, e.g., spine. And we evaluated our model on some open-source datasets, proving that our model trained on a large scale dataset has a better generalization performance or be great as a foundation for downstream tasks. 

\end{abstract}
\section{Introduction}
\begin{figure}[t]
	\centering
	\includegraphics[width=1.0\textwidth]{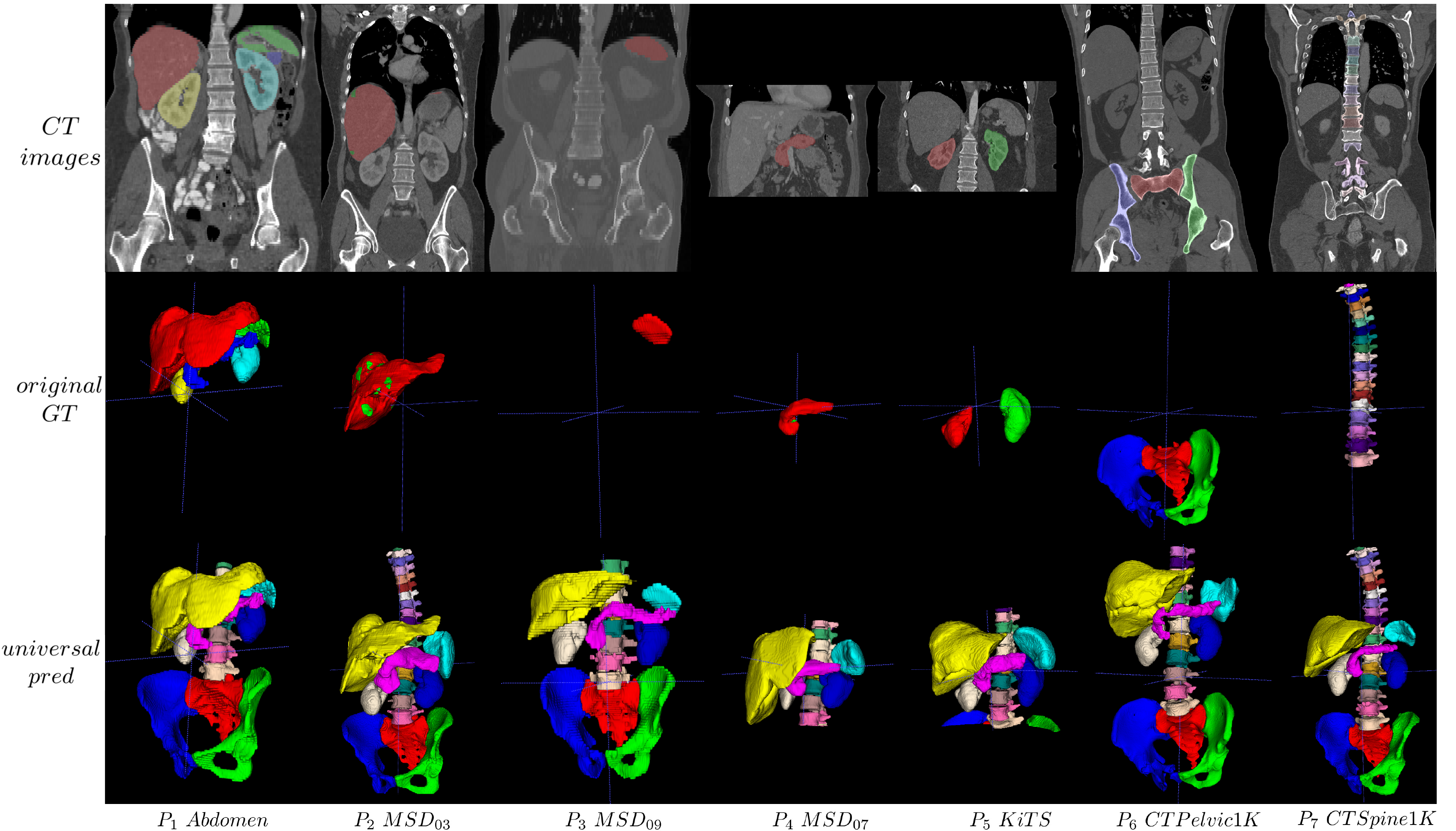}
	\caption{CT image examples of seven partially labeled datasets. 1$^{st}$ row shows the coronal view of CT images overlapped with original ground truth. 2$^{nd}$ row shows the 3D rendering of original ground truth. 3$^{rd}$ row shows the 3D rendering of results predicted from our 33-anatomy universal segmentation model.
	} \label{result_uni}
\end{figure}

Medical image segmentation~\cite{zhou2019handbook, zhou2021review} is a fundamental task in clinical workflow, which allows intelligent systems to know where the boundaries of target structures are. The more structures, the better. But it is practically hard to curate a large-scale, fully-annotated dataset, from which a model that segments myriad anatomies is learned, as annotating all anatomies of interest needs a support of professional knowledge from different doctors, and it is expensive and  time-consuming. Fortunately, there exists medical image datasets which are partially labeled according to the task at the time. As shown in the 1$^{st}$ and 2$^{nd}$ rows of Fig.~\ref{result_uni}, each dataset only contains the labels of a part of anatomies. Fusing these partially-labeled datasets together is a promising direction. Some methods~\cite{PIPO, PaNN, cvpr2021dodnet, ShiMargExc} have been proposed, verifying that such learned model outperforms the models that are individually trained from a single dataset. However, these methods deal with only abdominal organs with limited field-of-view (FOV).

On the other hand, although constructing open source datasets is a toilsome task, there are some recent datasets with a noticeable scale~\cite{flare21, pelvicSegDataset2021deep}. Ma \textit{et al.}~\cite{flare21} curate a multi-organ segmentation dataset of over 1,000 volumes from 12 sites with five abdominal organs completely labeled. The same goes for pelvic bone segmentation, too. Liu \textit{et al.}~\cite{pelvicSegDataset2021deep} construct a large-scale dataset containing over 1,000 volumes of pelvic bone structures. 

In this paper, we are interested in segmenting both the bones (spine and pelvis) and organs simultaneously. While there are sizeable organ datasets~\cite{MSD, kits19_url3} by now, there is a lack of sizeable bone datasets. Schnider \textit{et al.}~\cite{boneWholeBody20203d} segment 125 distinct bones in the upper-body CT, but with only 5 annotated cases. Pelvis and spine are important structures maintaining the stability of the body. The dataset gap in pelvic bone segmentation has been filled by Liu \textit{et al.}~\cite{pelvicSegDataset2021deep}, but the gap in vertebrae segmentation has not yet filled.
For vertebrae segmentation, the Verse challenge is a famous benchmark, but only 141 labeled cases for training~\cite{verse2021}. To this, we hereby curate images from many sources to construct a large-scale dataset, called \textbf{CTSpine1K}, containing 1,003 3D CT images and with over 11K vertebrae segmentation annotations. We will open source CTSpine1K.

Learning to segmenting the elongated spine and its associated vertebrae is challenged by the limited GPU memory. It is conventional to train a model using cropped patches as inputs instead a whole 3D volume, which limits the amount of contextual information to be learned. To this, we propose a \textbf{cross-patch transformer} module to fuse more information in adjacent patches, which enlarges the aggregated receptive field for improved segmentation performance. 

After CTSpine1K is built, we learn a universal model for segmentation of \textit{33 categories anatomies} from 7 partially-labeled datasets with about 2,800 volumes. The 33 anatomies include 3 pelvic bone, 5 abdomen organs, and 25 vertebrae. %An universal model which are compatible with multiple tasks meanwhile is a charming direction in artificial intelligence~\cite{universalZhuheqin2021you, universalLiuxinwen2021, universalLihan2021conditional, universalHuangchao20193d}.
We verify the generalization performance of our model on three open source datasets. We will make this universal model public too to benefit the community. %For organs segmentation, we can gain +2.4 Dice tested on Amos (is coming) directly. For vertebrae segmentation, we get +6.2 Dice on validation set of \textcolor{red}{Verse19} challenge after finetuning on the training set, even surpass the performance of the SOTA method~\cite{verse2021}, which include three stages in its workflow. But we are so efficient that we can get them in once inference. Using an unified segmentation model also has a well scalability in the future. And Liu et al.~\cite{incremental2021liu} verified the potential of class incremental learning in medical image scenario. [THIS PART can be left for experiment]

Our contributions can be summarized as:
\begin{itemize}
	\item We propose and will open source an \textit{universal model for segmentation of 33 anatomies} trained on 7 partially labeled datasets with about 2,800 volumes. Its generalization capability has been verified in other open source datasets.
	
	\item For vertebrae segmentation, we construct a large-scale CT dataset, \textit{CTSpine1K}, with 1,003 3D CT images and over 11K vertebrae segmentation annotations, to benefit the spine analysis community. 
	
	\item For spine segmentation of CTSpine1K, we also propose a \textit{cross-patch transformer module}~(CPTM) to catch more long-range contextual information to improve the accuracy of identification of vertebrae. 
\end{itemize}
%In the past, models for different tasks are all trained separately. It's not efficient in clinical usage, i.e., if we need many structures for the followed workflow, we need run many times of inference. And because of the limitation of the number of cases in the individual dataset, the performance of generalization of the model is also limited. 

%Reason; disadvantages. 
%-----------------
%Existing lots of partially labeled dataset. Training separated model for specific task in the past, with limited scale of dataset.
%
%\paragraph{Universal model}  We proposed an universal model for  segmentation of 33 categories anatomies trained on 7 partially labeled datasets [around 2500 volumes] via \textcolor{red}{pseudo label method. Will open source}.
%
%\paragraph{CTSpine1K} We proposed a new dataset for vertebrae segmentation task. \textcolor{red}{[??have been open sourced]}
%
%\paragraph{For long-range structure [Spine]} We proposed cross-patch transformer module to catch more information in longer distance to improve the accuracy of identification of vertebrae. Compared with the former SOTA method segmenting vertebrae via two phase framework, we segment the vertebrae directly in one step with other structures, which is more \textcolor{red}{practical?}
%
%Testing on Verse2019 \& Verse2020, we get SOTA results??
%Testing on Anhui Amos???
%
%Long training phase[10d+].

\section{Method}
\subsection{Cross-patch transformer}\label{cptm}
\begin{figure}[t]
	\centering
	\includegraphics[width=1.0\textwidth]{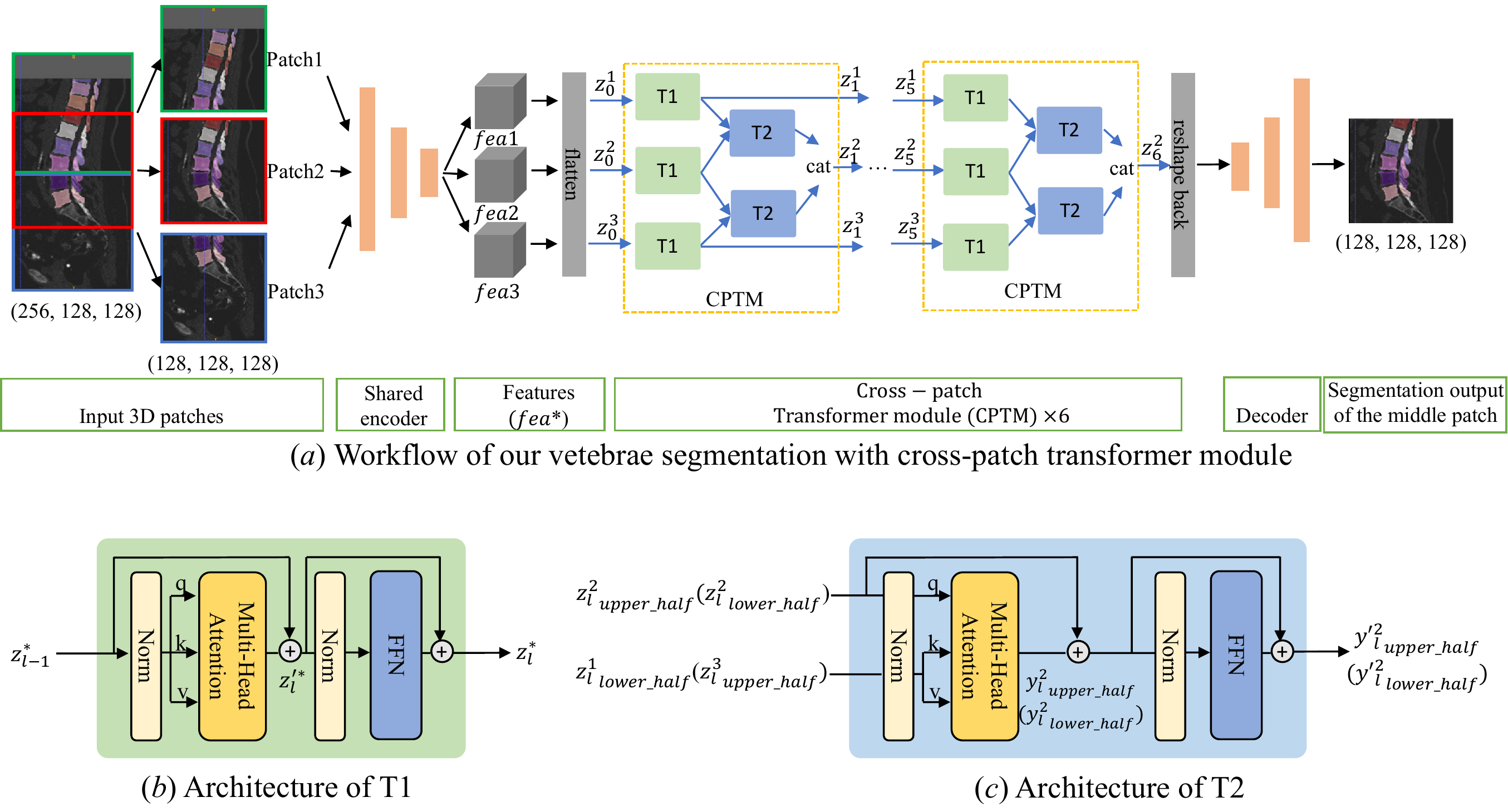}
	\caption{$(a)$ The overview of our spine segmentation workflow. The inputs are three consecutive 3D patches with overlapping. After being encoded by a shared encoder, the features of three patches are fused by a series of CPTMs. In each CPTM, $T_1$ and $T_2$ are used to extract information within one patch and fuse information between two patches, respectively. Then all contextual information is merged to the middle patch for final segmentation. $(b)$ and $(c)$ show the structure of the $T_1$ and $T_2$, respectively.
	} \label{cross_patchT}
\end{figure}

Due to the large volume size and limited GPU memory, patch-based method is a normal training paradigm in 3D medical image segmentation. This limits the receptive field of the model fundamentally. For example, when the patch size is $(128\times128\times128)$, the information can be caught is limited this patch size, regardless of the depth of the model. But for spine modeling,  long-range contextual information is important for identification of vertebrae. %, which are sequentially related. 
Although transformer~\cite{attentionisallyouneed2017attention} is famous at fusing long-range information in computer vision, there is no long-range information to catch in a single patch~\cite{transformer2021miccai3dAnatomyAware, transformer2022unetr}. 

We propose a cross-patch transformer module for modeling long-range context. As shown in Fig.~\ref{cross_patchT} $(a)$, we crop three patches consecutively with $(64\times128\times128)$ overlap as input. Then we can equivalently supply a `merged patch' with a double size of $(256\times128\times128)$ to the model. A shared encoder is used to extract information from each of three patches, respectively. Then the 
flattened feature maps are sent to a series of CPTMs for information fusion. All information of this double-sized `patch' is merged into the middle patch, which is decoded for the prediction of the segmentation.

In our CPTM, there are two kinds of transformer blocks, $T_1$ and $T_2$. $T_1$ follows a normal transformer mechanism, as shown in Fig.~\ref{cross_patchT} $(b)$ and Eq.~(\ref{T1}), fusing global information within a single input patch. %which will be mapped to $q, k, v$. 
Mathematically,
\begin{gather}\label{T1}
% \begin{align}
z'^*_l = MSA(LN(z^*_{l-1}))+z^*_{l-1},~~z^*_l = FFN(LN(z'^*_l))+z'^*_l,
% \end{align}
\end{gather}
where $z^*_{l-1}$, $z^*_l$ are input and output features, respectively. $MSA$ is multi-head attention, $LN$ is layer normalization, and $FFN$ is feed-forward network. 

$T_2$ is used to fuse information from two adjacent patches (the top `Patch1' and the bottom `Patch3') with that of the middle `Patch2'. The fusion proceeds as shown in Fig.~\ref{cross_patchT} $(c)$. First, it integrates the features of the intersection parts of two patches (say the lower half of `Patch1' and the upper half of `Patch2' or similarly the lower half of `Patch2' and the upper half of `Patch3') using Eq.~(\ref{T2}) to (\ref{T3_3}). Then, we concatenate the integrated features back to a  whole patch. This way, the information of three patches is merged together to the middle patch. 
%\textcolor{red}{we take the intersection part of two patches, respectively, as inputs, with each input of the half size of $T_1$'s input.} 
%Because the information has been fully integrated inside the patch by $T_1$, these two half-sized patch features contain all information of corresponding whole patch. After upper\_half and lower\_half of patch2 finish fusing information from adjacent patches, we concatenate them back to a whole patch. Through followed $T_1$ further integrating information in patch2, then three patches' information can be merged together to the middle patch. 
%\begin{gather}
\begin{align}\label{T2}
{y^2_l}_{upper\_half} &= MSA(LN({z^2_l}_{upper\_half}), LN({z^1_l}_{lower\_half}))+{z^2_l}_{upper\_half},\\
{y'^2_l}_{upper\_half} &= FFN(LN({y^2_l}_{upper\_half}))+{y^2_l}_{upper\_half},\\
z^2_l &= torch.cat({y'^2_l}_{upper\_half}, {y'^2_l}_{lower\_half}). \label{T3_3}
\end{align}
%\end{gather}

%Almost all existing 3D segmentation methods are designed within one patch. 

\subsection{Pseudo label prediction and universal model learning}

%From each dataset with partial label, we first train a model for these labeled organs; this way we have multiple models, each segmenting some anatomies. Then we run these models on all training images in all datasets to generate pseudo labels. 

Learning from predicted pseudo labels is a simple but efficient semi-supervised method. Because the structures of humans are similar, in addition to the annotations that already exist in the partially labeled datasets, there also exists \textit{a large amount of unlabeled anatomies} in images waiting to be mined and utilized.

%This characteristic of medical image gives us an intuitive insight, that is, we can learn a universal segmentation model simply via mining already existing organ labels and pseudo labels generated based on large amount of partially labeled datasets. 

To construct a universal model to segment organs and bones, we curate 7 partially labeled datasets, and train three models separated based on the SOTA methods for organs~\cite{ShiMargExc}, pelvic bones~\cite{FabianNNUnet_nm}, and vertebrae (with CPTM in \ref{cptm}), respectively. Then we predict all 33 anatomies for all images in these 7 partially labeled datasets as pseudo labels, which are replaced by original ground truth labels if present, constructing a 33 classes `fully labeled' segmentation dataset. Finally, we train a 33-anatomy segmentation model using nnU-Net~\cite{FabianNNUnet_nm}.

\section{Experiments}
\begin{table}[t]
	\centering
	\caption{A summary of 7 training \& validation  datasets and 4 independent testing datasets used in our experiments.}\label{summaryofdatasets}
	\newsavebox{\tablebox}
	\begin{lrbox}{\tablebox}
		\begin{tabular}{c|l|c|c|l|c|l}
			\thickhline
			Phase&Datasets &  Modality &  \# of labeled volumes & Annotated organs & Mean spacing (z, y, x) & Source\\
			\hline\hline
			\multirow{8}{2cm}{\centering $Training$\\$\&$\\$Validation$}&Dataset1 ($P_1$) 	 & CT & 30 	& Five organs 	& (0.76, 0.76, 3.0) &Abdomen in \cite{matlas}\\
			&Dataset2 ($P_2$) & CT & 131 & Liver 		& (0.77, 0.77, 1.0) 	&MSD\_Liver~\cite{MSD}\\
			&Dataset3 ($P_3$) & CT & 41 	& Spleen 			& (0.79, 0.79, 1.6) &MSD\_Spleen~\cite{MSD}\\
			&Dataset4 ($P_4$) & CT & 281 & Pancreas 	& (0.80, 0.80, 2.5) 		&MSD\_Pancreas~\cite{MSD}\\
			&Dataset5 ($P_5$) & CT & 210 & L\&R Kidneys & (0.78, 0.78, 0.8) 		&KiTS~\cite{kits19_url3}\\
			&Dataset6 ($P_6$) 	 & CT & 1109 	& Pelvic Bones 	& (0.78, 0.78, 1.5) &CTPelvic1K \cite{matlas}\\
			&Dataset7 ($P_7$) 	 & CT & 1005  	& Vertebrae 	& (0.76, 0.76, 1.1) &CTSpine1K($Ours$)\\
			\cline{2-7}	
			&All				 & CT & 2807 & Bones \& organs & (0.77, 0.77, 1.4) & -\\
			\hline\hline
			\multirow{4}{2cm}{\centering $Testing$}&Amos				 & CT & 200 & Five organs &(0.74, 0.74, 5.0)  & Coming soon\\
		    &CLINIC				 & CT & 200 & Five organs &(0.74, 0.74, 1.2)  & In-house\\
			&Verse19				 & CT & 80/40/40 & Vertebrae & (1.00, 1.00, 1.6)  & Challenge~\cite{verse2021}\\
			&Verse20				 & CT & 120/103/103 &Vertebrae&(0.80, 0.80, 1.4)  & Challenge~\cite{verse2021}\\
			\thickhline
		\end{tabular}
	\end{lrbox}
	\scalebox{0.7}[0.7]{\usebox{\tablebox}}
\end{table}

\subsection{Datasets}
In this part, we introduce the datasets used in our experiments, including CTSpine1K we construct. Table~\ref{summaryofdatasets} provides a summary of these datasets.

%\subsubsection{\textcolor{red}{CTSpine1K}}

\subsubsection{CTSpine1K Dataset}
To build a comprehensive spine dataset that replicates practical appearance variations, we curate a large-scale dataset of CT images that contain spinal vertebrae from the following four open sources: COLONOG~\cite{johnson2008accuracy}, HNSCC-3DCT-RT~\cite{bejarano9head}, MSD T10~\cite{MSD}, and COVID-19~\cite{harmon2020artificial}.

We reformat all DICOM images to NIfTI to simplify data processing and de-identify images to meet the institutional review board (IRB) policies of contributing sites. All existing sub-datasets are under Creative Commons license CC-BY-NC-SA and we will keep the license unchanged. For sub-dataset MSD T10 and sub-dataset COVID-19, we choose some cases from them, and in all these data sources, we exclude those cases of very low quality. The overview of our dataset can be seen in Table~\ref{tab1}. The details about the CTSpine1K dataset and the annotation pipeline could be seen in supplementary material. 

    \begin{table}[t]
    \centering
    \caption{Overview of our large-scale CTSpine1K dataset. Ticks[\checkmark] in the table refer to that this dataset contains some special cases with sacral lumbarization or lumbar sacralization and we will list their IDs in the open source. We exclude the metal-artifact cases due to the difficulty of labeling them.}\label{tab1}
    \resizebox{\linewidth}{!}{
    \begin{tabular}{l|c|c|c|c|c}

    %\toprule[1pt].
    \thickhline
    Dataset name &Cases & \# of vertebrae & Mean spacing(mm) & Mean size &Source and Year\\
    % \hline

    % VerSe'19,20 [\checkmark] & 41 &\makecell[c]{Cervical, thoracic and \\lumbar vertebrae (501)}&(0.81, 0.81, 0.88)&(512, 512, 640)&URL\footnotemark[1]$^,$\footnotemark[2] 2019, 2020\\
    \hline
    \hline
    COLONOG [\checkmark]&  784 &\makecell[c] {Thoracic and lumbar \\vertebrae (8,136)}&  (0.75, 0.75, 0.8)&(512, 512, 542)& \cite{johnson2008accuracy} 2008\\
    HNSCC-3DCT-RT& 31 &\makecell[c]{Cervical and thoracic \\vertebrae (450)}& (1.09, 1.09, 2.0)&(512, 512, 202)& \cite{bejarano9head} 2018\\
    MSD T10 [\checkmark] & 148 &\makecell[c]{Thoracic and lumbar \\vertebrae (2,101)}&  (0.78, 0.78, 1.6)&(512, 512, 458)& \cite{MSD} 2019\\
    COVID-19 & 40 &\makecell[c]{Cervical and thoracic \\vertebrae (612)}&  (0.79, 0.79, 4.5)&(512, 512, 93)& \cite{harmon2020artificial} 2020\\
    \hline
    CTSpine1K [\checkmark]& 1,003 &\makecell[c]{Cervical, thoracic and \\lumbar vertebrae (11,299)}&(0.77, 0.77, 1.1)&(512, 512, 501)&-\\
    %\toprule[1pt]
    \thickhline

    \end{tabular}
    }
    \end{table}
    % \footnotetext[1]{https://verse2019.grand-challenge.org/}
    % \footnotetext[2]{https://verse2020.grand-challenge.org/}

\subsubsection{Other datasets} The other 6 partially labeled datasets in training phase are: CTPelvic1K~\cite{pelvicSegDataset2021deep}, MSD\_Liver~\cite{MSD}, MSD\_Spleen~\cite{MSD}, MSD\_Pancreas~\cite{MSD}, KiTS~\cite{kits19_url3}, and Abdomen~\cite{matlas}. The details are shown in Table~\ref{summaryofdatasets}. Different from CTSpine1K, we split these 6 datasets into training: testing with ratio of $4:1$, respectively. For fair model selection, we choice the model saved in the last epoch.

We also leverage 4 independent testing datasets to verify the generalization capacity of our universal model: Amos, CLINIC, Verse19~\cite{verse2021}, and Verse20~\cite{verse2021}. Amos is temporarily private, and will open source soon. CLINIC is an in-house dataset by now.

%\subsection{How to finetune to other datasets}
%marginal loss finetuning vs finetuning directly???
%Based on our universal model, we can implement it into our actual production environment directly, but also can treat it as a pretrained model.  Usually, we only need to finetune part of 33 classes. If we finetune it directly with our own new `partially labeled' dataset, the existing `knowledge conflict' will slow our finetuning, 

\subsection{Results and discussion}
\subsubsection{Evaluation metrics}\label{evaluationmetric}
To evaluate the performance of different methods, we employ widely-used segmentation metrics, including \textit{Dice coefficient (DC)} and \textit{Hausdorff distance (HD)}. 
For organs evaluation, we use the $95^{th}$ percentile HD (HD95) to measure the degree of false positive prediction. Due to the computing pressure and refering to challenge Verse19 and Verse20, for bones evaluation, we use HD instead, equiped with maximum connected region post-processing to prevent influence of outlier. In addition, to evaluate the performance of vertebrae localization~\cite{verse2021}, we also compute the \textit{Identification Rate (id.rate)} and \textit{Localization distance ($d_{mean}$)}. Because there is no landmark detection output from our segmentation model, we define the centroid of each vertebra as a landmark. %Although the landmark of different vertebrae under the new standard are not uniform, the landmark of same vertebra is unified between different patients. 

  \begin{table}[t]
 	\centering
 	\caption{The average DC and HD results for different methods, validated on different partially labeled datasets. [S] means separately trained model for specific task, i.e., pelvic bone, organs, and vertebrae segmentation. [Uni] means an universal model trained for 33 categories via pseudo label method with all partially labeled datasets. The number in [\ ] means how many classes in that structure. $[(6)]$ and $[(25)]$ mean that $\#$25 vertebra is included. `-' means no result to show. }\label{mainresults}
 	\begin{lrbox}{\tablebox}
 		\begin{tabular}{l|rr|rr||rr|rr|rr||rr}
 			\thickhline
 			\multirow{2}{3cm}{\centering Structures\\$[\# of classes]$} &  \multicolumn{2}{c}{[S]nnU-Net~\cite{FabianNNUnet_nm}} &   \multicolumn{2}{c||}{[S]MargExc~\cite{ShiMargExc}} &  \multicolumn{2}{c}{[S]nnFormer~\cite{nnformer2021nnformer}}  &  \multicolumn{2}{c}{[S]nnU-Net~\cite{FabianNNUnet_nm}} &  \multicolumn{2}{c||}{[S]CPTM}&  \multicolumn{2}{c}{[Uni] Ours} \\
 			\cline{2-13}
 			&  DC& HD&  DC&HD &DC& HD &DC&HD & DC&HD& DC& HD\\
 			\hline\hline
 			%				$Sacrum$ & .965 & 7.867&-&-& 	-	&-&-&-&-  	&-&.966&8.320\\
%				$L\ Hip$ & .973 & 4.424 &- 		& -&- 	&-&-&-&-  	&-&.977&4.807\\
%				$R\ Hip$ & .975 &5.037  &- 		&-&-  	&-&-&-&-  	&-&.977&5.576\\
 			Pelvic bones mean $[3]$& .971 & \textbf{5.78} &- 		&-&-  	&-&-&-&-  	&-&\textbf{.973}&6.23\\
 			\hline
 			%				$Liver$ & - &-&.962  		& 7.010 &-	&-&-&-& - &-&.968&2.898\\
%				$Spleen$ 	 & - &-&.965   		&1.148 &- &-	&-&-& - &-&.965&1.148\\
%				$Pancreas$ 	 & - &-&.848   		&4.831&- &-	&-&-&-  &-&.848&5.210\\
%				$R\ Kidney$ 	& - &-&.968   	&1.390	&-&- &- 	&-&-  &-&.970&1.239\\
%				$L\ Kidney$ 	 & - &-&.965   	&3.955&-& - 	&-&-&-  &-&.958&2.267\\
%				\hline
  			Organs mean $[5]$ & - &-  &.942		&3.67 &-  	&-&-&-&-  	&-&\textbf{.942}&\textbf{2.55}\\
 			\hline
 			Cervical V $[7]$ 	 & - &-  & -&-	&.626	&9.94& \textcolor{red}{.816} 	&\textcolor{red}{7.76}& .807 	&8.38&\textbf{.831}&\textbf{7.19}\\
 			Thoracic V $[12]$ 	& - &-  &- & -&	.806	&12.44& .870 	&9.97&  \textcolor{red}{.880}	&\textcolor{red}{9.68}&\textbf{.884}&\textbf{9.38}\\
 			Lumbar V $[5]$ 	& - & - &- &- 	&.906 &11.72 & .933	&8.36&\textbf{.935}	&\textcolor{red}{8.26}&\textcolor{red}{.934}&\textbf{8.18}\\
 			Lumbar V $[(6)]$ 	& - & - &- &- 	&.755	& 11.72& .816 &\textcolor{red}{8.59}&\textbf{.818}  	&\textbf{8.41}&\textcolor{red}{.817}&11.65\\
 			\hline
 			All V mean $[24]$ & - & - & -	&-&.774	&11.56 &.867 &\textcolor{red}{8.99} & \textcolor{red}{.870} & 9.01 & \textbf{.879} & \textbf{8.49}\\
 			All V mean $[(25)]$ & - & - & -	&-&.743	&11.56  &.842&\textbf{9.02}&\textcolor{red}{.845}  	&\textcolor{red}{9.03}&\textbf{.853}&9.31\\
 			\thickhline
 		\end{tabular}
 	\end{lrbox}
 	\scalebox{0.78}[0.78]{\usebox{\tablebox}}
 \end{table}

 \begin{table}[t]
	\centering
	\caption{The performance of landmark localization of different segmentation models under the newly defined landmark~(\ref{evaluationmetric}). HNSCC-3DCT-RT, MSD\ T10, COVID-19, and COLONOG are different sources in CTSpine1K dataset. $id.rate$ is reported in \% and $d_{mean}$ in mm. }\label{IdrateDist}
	\begin{lrbox}{\tablebox}
		\begin{tabular}{l|c|rr|rr|rr||rr}
			\thickhline
			\multirow{2}{1.5cm}{\centering Vertebrae\\source}&\multirow{2}{0.9cm}{\centering \# of Scans}  & \multicolumn{2}{c}{[S]nnFormer~\cite{nnformer2021nnformer}} & \multicolumn{2}{c}{[S]nnU-Net~\cite{FabianNNUnet_nm}} & \multicolumn{2}{c||}{[S]CPTM} & \multicolumn{2}{c}{[Uni]nnU-Net} \\
			\cline{3-10}
			&& $id.rate$ &$d_{mean}$&$id.rate$ &$d_{mean}$&$id.rate$ &$d_{mean}$&$id.rate$ &$d_{mean}$\\
			\hline\hline
			HNSCC-3DCT-RT		& 5+5	&\textcolor{red}{98.38} &1.93  & 	98.18	&1.37&  \textbf{100}	&\textcolor{red}{0.87}	&\textbf{100}&\textbf{0.67}\\
			MSD\ T10	&  30+30&95.44  &2.11  & 97.05		&1.42&\textbf{97.90}  	&\textbf{1.21}	&\textcolor{red}{97.31}&\textcolor{red}{1.29}\\
			COVID-19 	& 10+10	& 94.60 &2.75  & \textcolor{red}{98.28}	&\textcolor{red}{1.22}&\textbf{98.92}  	&\textbf{1.01}	&97.23&1.49\\
			COLONOG	& 152+152	& \textcolor{red}{96.80} &1.46  & 96.61	&1.38&96.51  	&\textcolor{red}{1.36}	&\textbf{97.18}&\textbf{1.19}\\
			\hline
			All		& 197+197	&96.52  &1.64  & 96.80	&1.37&\textcolor{red}{96.90}  	&\textcolor{red}{1.31}   &\textbf{97.27}&\textbf{1.20}\\
			\thickhline
		\end{tabular}
	\end{lrbox}
	\scalebox{0.92}[0.92]{\usebox{\tablebox}}
\end{table}

\subsubsection{Cross-patch transformer module (CPTM)} 
Validation results of models trained separately for different part of anatomies are shown in Table~\ref{mainresults}. For vertebrae segmentation in CTSpine1K, we implement current SOTA CNN-based and transformer-based methods~\cite{FabianNNUnet_nm, nnformer2021nnformer} to find a more powerful model to supply more precise pseudo label for next stage of universal model training. We find that nnU-Net can keep more stable performance in different scenarios, so we deploy our CPTM into nnU-Net framework. There is only {slight improvement on DC} compared with nnU-Net, but in Table~\ref*{IdrateDist} we find CPTM can improve {vertebrae identification rate obviously}.  This means the CPTM can fuse more useful context to enhance the performance of segmentation. Experiment of nnU-Net trained on same size input of CPTM  is not implemented because of the limit of GPU memory.

\subsubsection{Universal model}
Considering the wide range of applications of nnU-Net~\cite{FabianNNUnet_nm}, we train our universal model based on it. On validation set, as shown in Table~\ref{mainresults} and Table~\ref{IdrateDist}, there is an {obvious} improvement on vertebrae segmentation. For pelvic bones and five organs segmentation, existing methods~\cite{FabianNNUnet_nm, ShiMargExc} have almost reached the upper bound performance, so there is {not very obvious improvements on pelvic bones and organs segmentation}. 

More importantly, we verify the generalization capacity of our universal model on other testing datasets, which are never seen in the training phase. As shown in Table~\ref{generalization}, when we deploy our universal model directly for testing organs segmentation, we can achieve the improvement of 2.5 and 2.0 in Dice and  of 35.7\% and 23.5\% in {HD95} on Amos and CLINIC datasets, respectively. For vertebrae segmentation, we choose two famous challenges~\cite{verse2021}, Verse19 and Verse20, for testing our model's generalization performance. {In Table~\ref{generalization}, to compare with SOTA methods in these two challenges, we modify the evaluation metric of DC and HD from vertebra-level to patient-level, referring to~\cite{verse2021}, without post-processing.} When we directly test on validation and testing set of Verse19 and Verse20, the performance on Verse19 is not good compared with model trained from scratch. But after finetuning on training set, our model's performance gets a big improvement. The performance on verse19 surpasses the SOTA method, which have a complicated three stage design~\cite{verse2021}. The performance on verse20 also gets a big boost after finetuning on training dataset. For pelvic bone segmentation, there is no large scale open source datasets to test.

For the visualization of segmentation results of our universal model, please refer to the 3$^{rd}$ row in Fig.~\ref{result_uni}. The 33 anatomies can be predicted by only one inference, a feat never realized before.

\begin{table}[t]
	\centering
	\caption{Generalization performance of our universal model on other testing datasets. `f.t.' means `finetuned to'. `f.t.s' means `from the scratch'. $id.rate$ is reported in \% and $d_{mean}$ in mm. Attention: in SOTA method of~\cite{verse2021}, the landmark definition are different.}\label{generalization}
	\begin{lrbox}{\tablebox}
		\begin{tabular}{l|r|rr|rr||rr|rr||rr|rr||rr|rr}
			% 			\cline{0-13}
			\thickhline
			\multirow{2}{1.5cm}{\centering Vertebrae}&\multirow{2}{0.9cm}{\centering \# of Scans}& \multicolumn{4}{c}{[S]MargExc~\cite{ShiMargExc}} & \multicolumn{4}{c}{[S]Verse19 f.t.s~\cite{FabianNNUnet_nm}} & \multicolumn{4}{c}{[S]Verse20 f.t.s~\cite{FabianNNUnet_nm}} &\multicolumn{4}{c}{}\\
			\cline{3-14}
			&&DC &HD&$id.rate$ &$d_{mean}$&DC&HD&$id.rate$ &$d_{mean}$&DC&HD&$id.rate$ &$d_{mean}$&\multicolumn{4}{c}{}\\
			\cline{0-13}
			Amos		& 200	& .855 & 10.20 & -&-&- &-&-&-&-&-&-&-&\multicolumn{4}{c}{}\\
			CLINIC	& 107&.934  &3.23  & -&-&-  	&-&-&-&-&-&-&-&\multicolumn{4}{c}{}\\
			\cline{0-13}
			Verse19\ val 	& 40	& - &-  & -	&-&  .857	  &14.95&92.9&3.11&-&-&-	&-&\multicolumn{4}{c}{}\\
			Verse19\ test 	& 40	& - &-  & -	&-&.863	&27.21&92.9&3.16&-&-&- &-&\multicolumn{4}{c}{}\\
			\cline{0-13}
			Verse20\ val 	& 103	& - &-  &- 	&-& -  &-&-&-&.805&34.38&87.9 &4.72&\multicolumn{4}{c}{}\\
			Verse20\ test 	& 103	&-  &-  &- 	&-& - 	  &-&-&-&.829&29.07&89.9&3.98&	\multicolumn{4}{c}{}\\
			\hline\hline
			\multirow{2}{1.5cm}{\centering Vertebrae}&\multirow{2}{0.9cm}{\centering \# of Scans} & \multicolumn{4}{c}{[Uni]nnU-Net} & \multicolumn{4}{c}{[Uni] f.t. Verse19} & \multicolumn{4}{c||}{[Uni] f.t. Verse20}&	\multicolumn{4}{c}{SOTA in ~\cite{verse2021}}\\
			\cline{3-18}
			&&DC &HD&$id.rate$ &$d_{mean}$&DC&HD&$id.rate$ &$d_{mean}$&DC&HD&$id.rate$ &$d_{mean}$&DC&HD&$id.rate$ &$d_{mean}$\\
			\hline
			Amos		& 200	& \textbf{.880} & \textbf{6.56 }& 	-	&-&  -	&-	&-&-&-&-&-&-&-&-&-&-\\
			CLINIC	& 107&\textbf{.954}  &\textbf{2.47}  & 	-	&-& -	&-	&-&-&-&-&-&-&-&-&-&-\\
			\hline
			Verse19\ val 	& 40	& .808 & 18.64 &89.0 	&3.31& \textbf{.914}&\textbf{9.40}&\textbf{99.7}&\textbf{0.91}&-&-&-&-&.909&6.35&95.7&4.27\\
			Verse19\ test 	& 40	&.829  &17.08  &93.4 	&3.56&\textbf{.904} &\textbf{10.28}&\textbf{96.8}&\textbf{1.55}&-&-&-	&-&.898&7.08&94.3&4.80\\
			\hline
			Verse20\ val 	& 103	&.797  &32.08  &87.5 	&4.69&  - &-&-&-&\textbf{.834}&\textbf{18.41}&\textbf{89.4}&\textbf{4.01}&.917&5.80&95.1&2.90\\
			Verse20\ test 	& 103	&.824  &26.40  &91.5 	&3.54&- &-&-&-&\textbf{.866}&\textbf{18.03}&\textbf{93.2}	&\textbf{2.59}&.897&6.06&92.8&2.91\\
			\thickhline
		\end{tabular}
	\end{lrbox}
	\scalebox{0.69}[0.69]{\usebox{\tablebox}}
\end{table}

\section{Conclusion}

In this paper, to solve the problem that a large-scale, fully-labeled dataset in medical image segmentation is difficult to obtain then a robust model cannot reliably be trained, we fuse 7 partially labeled datasets based on pseudo label method to train a 33-class segmentation model. Because the structure of the human body is similar, our method can effectively unleash the potential of a large amount of unlabeled data in partially labeled datasets. 
This universal 33-class segmentation model has good generalization performance and can also be used as a pretrained model to benefit subsequent tasks.
Among them, a large-scale vertebrae segmentation dataset is constructed by us, CTSpine1K, containing 1,003 3D CT images and over 10k vertebrae annotations, which effectively fill the gap of data insufficiency in spine analysis. To account for the elongated nature of a spine, we also propose a cross-patch transformer module (CPTM) to improve the accuracy of vertebrae identification. %How to use contextual information more effectively is a meaningful research direction in 3D medical image segmentation. 
Both the 33-anatomy universal segmentation model and the CTSpine1K dataset will be open source. We plan to keep building a universal model to include more anatomies in future.

 \bibliographystyle{splncs04}
 \bibliography{Manuscript_PseudoLabel}
%
%\begin{thebibliography}{8}
%\bibitem{ref_article1}
%Author, F.: Article title. Journal \textbf{2}(5), 99--110 (2016)
%
%\bibitem{ref_lncs1}
%Author, F., Author, S.: Title of a proceedings paper. In: Editor,
%F., Editor, S. (eds.) CONFERENCE 2016, LNCS, vol. 9999, pp. 1--13.
%Springer, Heidelberg (2016). \doi{10.10007/1234567890}
%
%\bibitem{ref_book1}
%Author, F., Author, S., Author, T.: Book title. 2nd edn. Publisher,
%Location (1999)
%
%\bibitem{ref_proc1}
%Author, A.-B.: Contribution title. In: 9th International Proceedings
%on Proceedings, pp. 1--2. Publisher, Location (2010)
%
%\bibitem{ref_url1}
%LNCS Homepage, \url{http://www.springer.com/lncs}. Last accessed 4
%Oct 2017
%\end{thebibliography}

\end{document}